# The Spontaneous Emergence of Conventions: An Experimental Study of Cultural Evolution

Damon Centola[a,b] and Andrea Baronchelli[c,*]

a. Annenberg School of Communication, University of Pennsylvania, Philadelphia, PA 19106
b. Center for Advanced Study in the Behavioral Sciences, Stanford University, Palo Alto, CA 94305
c. Department of Mathematics, City University London, Northampton Square, London EC1V 0HB, UK

**Abstract:**

How do shared conventions emerge in complex decentralized social systems? This question engages fields as diverse as linguistics, sociology and cognitive science. Previous empirical attempts to solve this puzzle all presuppose that formal or informal institutions, such as incentives for global agreement, coordinated leadership, or aggregated information about the population, are needed to facilitate a solution. Evolutionary theories of social conventions, by contrast, hypothesize that such institutions are not necessary in order for social conventions to form. However, empirical tests of this hypothesis have been hindered by the difficulties of evaluating the real-time creation of new collective behaviors in large decentralized populations. Here, we present experimental results – replicated at several scales – that demonstrate the spontaneous creation of universally adopted social conventions, and show how simple changes in a population's network structure can direct the dynamics of norm formation, driving human populations with no ambition for large scale coordination to rapidly evolve shared social conventions.

* To whom correspondence should be addressed. Email: a.baronchelli.work@gmail.com.

**Significance Statement:**

Social conventions shape every aspect of our lives, from how we greet each other to the languages we speak. Yet, their origins have been a topic of theoretical speculation since the time of Aristotle. Most approaches assume that institutions are necessary to organize large populations, but the simplest explanation is that universally accepted conventions are the unintended consequence of individuals' efforts to coordinate locally with one another. While this hypothesis is compelling, it lacks conclusive empirical support. Here, we present results from controlled experiments demonstrating that changes in network connectivity can cause global social conventions to spontaneously emerge from local interactions, even though people have no knowledge about the population, or that they are coordinating at a global scale.

**Introduction**

Social conventions are the foundation for social and economic life (1-7), Yet, it remains a central question in the social, behavioral, and cognitive sciences to understand how these patterns of collective behavior can emerge from seemingly arbitrary initial conditions (2-4, 8, 9). Large populations frequently manage to coordinate on shared conventions despite a continuously evolving stream of alternatives to choose from, and no *a priori* differences in the expected value of the options (1, 3, 4, 10). For instance, populations are able to produce linguistic conventions on accepted names for children and pets (11), on common names for colors (12), and on popular terms for novel cultural artifacts, such as referring to junk email as "SPAM" (13, 14). Similarly, economic conventions, such as bartering systems (2), beliefs about fairness (3), and consensus regarding the exchangeability of goods and services (15), emerge with clear and widespread agreement within economic communities, yet vary broadly across them (3, 16).

Prominent theories of social conventions suggest that institutional mechanisms – such as centralized authority (14), incentives for collective agreement (15), social leadership (16), or aggregated information (17) – can explain global coordination. However, these theories do not explain whether, or how, it is possible for conventions to emerge when social institutions are not already in place to guide the process. A compelling alternative approach comes from theories of social evolution (2, 18-20). Social evolutionary theories maintain that networks of locally interacting individuals can spontaneously self-organize to produce global coordination(21, 22). While there is widespread interest in this approach to social norms (6, 7, 14, 18, 23-26), the complexity of the social process has prevented systematic empirical insight into the thesis that these local dynamics are sufficient to explain universally adopted conventions (27, 28).

Several difficulties have limited prior empirical research in this area. The most notable of these limitations is scale. While compelling experiments have successfully shown the creation of new social conventions in dyadic and small group interactions (29-31) the results in small group settings can be qualitatively different from the dynamics in larger groups (see SI text), indicating that small group experiments are insufficient for demonstrating whether or how new conventions endogenously form in larger populations (32, 33). Important progress on this issue has been made using network-based laboratory experiments on larger groups (15, 24). However this research has been restricted to studying coordination among players presented with two or three options with known payoffs. Natural convention formation, by contrast, is significantly complicated by the capacity of individuals to continuously innovate, which endogenously expands the 'ecology' of alternatives under evaluation (23, 29, 31). Moreover, prior experimental studies have typically assumed the existence of either an explicit reward for universal coordination (15), or a mechanism that aggregates and reports the collective state of the population (17, 24), which has made it impossible to evaluate the hypothesis that global coordination is the result of purely local incentives.

More recently, Data Science approaches to studying norms have addressed many of these issues by analyzing behavior change in large online networks (34). However, these observational studies are limited by familiar problems of identification that arise from the inability to eliminate the confounding influences of institutional mechanisms. As a result, previous empirical research has been unable to identify the collective dynamics through which social conventions can spontaneously emerge (8, 34-36).

We addressed these issues by adopting a Web-based experimental approach. We studied the effects of social network structure on the spontaneous evolution of social conventions in populations without any resources to facilitate global coordination (9, 37).  Participants in our study were rewarded for coordinating locally, however they had neither incentives, nor information for achieving large scale agreement.  Further, to eliminate any pre-existing bias in the evolutionary process, we studied the emergence of arbitrary linguistic conventions, in which none of the options had any *a priori* value or advantage over the others (3, 23).  In particular, we considered the prototypical problem of whether purely local interactions can trigger the emergence of a universal naming convention (38, 39).

**Theoretical model**

The approach used here builds on the general model of linguistic conventions proposed by Wittgenstein (39), in which repeated interaction produces collective agreement among a pair of players.   Theoretical extensions of this approach have argued that myopic players interacting in social networks can unintentionally create percolating cascades of coordinated behavior (6, 10, 23, 25, 27, 40, 41).  Theoretical predictions for our study are based on a derived 'language game' model of convention formation (27), in which agents attempting to coordinate in pairwise interactions accrue a memory of past plays, which they use to "guess" the words that will be used by their subsequent partners. Consistent with a broad range of formal approaches (5, 33, 42-44), this model predicts that the connectivity of the actors' social networks can influence the collective dynamics of convention formation, ranging from the emergence of competing regional norms that inhibit global coordination (45), to the rapid growth of universally shared social conventions (27) (see SI text).

We evaluated these predictions by studying convention formation in three representative network configurations:  *i*) spatially embedded social topologies (i.e., one dimensional lattices with degree 4)(45-47), *ii*) randomly connected topologies (i.e., random graphs with constant degree 4)(42, 48), and *iii*) homogeneously mixing populations (3, 27). Formal results show that alternative network configurations (48) fall within the range of dynamical behavior exhibited by the three topologies used here (42).

**Experimental Design**

Each live game, or experimental 'trial,' consisted of a set of participants, a specific social network structure, and a pre-specified number of rounds to play. When participants arrived to play the game, they were randomly assigned to positions within a social network. In a given round of the game, two network 'neighbors' were chosen at random to play with one another. Both players simultaneously assigned names to a pictured object (i.e., a human face), blindly attempting to coordinate in the real-time exchange of naming choices (see SI text). If the players coordinated on a name, they were rewarded with a successful payment; if they failed, they were penalized (see Materials and Methods). After a single round, the participants could see only the choices that they and their partner had made, and their cumulative pay was updated accordingly. They were then randomly assigned to play with a new neighbor in their social network, and a new round would begin. The object that participants were trying to name was the same for the entire duration of the game, and for all members of the game. Participants in the study did not have any information about the size of the population that was attempting to coordinate, nor about the number of neighbors that they were connected to.

**Results**

Figure 1 ($N$=24) shows that the dynamics of emergent social conventions depend decisively upon the structure of the social network. In spatial networks, populations enjoyed rapid local coordination, in some cases achieving a 50% success rate (i.e., average likelihood of matching words with a partner) as early as round 4. However, this initial success rate quickly decelerated. After 25 rounds of play, average success rates failed to reach above 75%. Throughout all of the spatial network trials, the dominant local conventions (i.e., the most popular word choices) were never used by more than 45% of the population. As shown in Figure 2, behavior in the spatial networks evolved through a process of local coarsening (45), in which emergent regions of coordinated behavior competed with bordering local conventions (45, 49, 50). In each of the trials, these dynamics inhibited the spontaneous emergence of global coordination by creating entrenched competition between endogenously formed groups (51).

Similar results were found in random networks, in which local groups of coordinated individuals emerged and competed for dominance (fig.1). Group formation in random networks was driven by repeated interactions, which created metastable boundaries between groups of neighbors despite the absence of local clustering (fig.2). After 25 rounds of play in randomized topologies, local groups still persisted and coordination rates never increased above 75%. In all random network trials, global social conventions never emerged. Moreover, the sizes of the dominant social conventions (i.e., the fraction of the population using the most popular word choice) were equivalent across all trials of the spatial and random networks, averaging 33% of the population. On timescales observable within our study, the dynamics of social coordination in both the spatial and random network trials were driven by local group competition, which impeded the emergence of global conventions.

Homogeneously mixing populations exhibited significantly different dynamics than those observed in the other two topologies. Initially, success rates were lower because actors did not have repeated interactions with their partners, which prevented "neighborhoods" of entrenched behavior from forming. However, local failure accelerated global coordination. In all trials with homogenously mixing populations, success rates increased to 100% well before the end of the study. Figure 1 shows that this rapid growth in individual success corresponds to the spontaneous emergence of a global social convention. In all trials, an emergent convention grew quickly, reaching over 60% of the population by Round 12, and achieving universal adoption between Rounds 20 and 22. On average, by Round 22 players who had never interacted with one another were all using the same convention, and were able to consistently coordinate with new partners.

The speed of self-organized conventions in these networks raises the question of whether these coordination dynamics scale up as population sizes increase. There are good reasons for skepticism. As system size increases, so does the expected number of competing alternatives circulating in the population; at the same time, because interactions are limited by the number of rounds in the game, increasing system size reduces the fraction of the population with which any given individual can interact. These considerations suggest that global coordination may be much more difficult in larger populations.

We tested this conjecture by doubling the size of the population and replicating our study. These larger trials ($N$=48) permit a more detailed view of the evolving competitive landscape that constitutes the 'ecology' of social conventions. Figure 3 shows the changing distribution of popularity among the competing alternatives in all three network conditions, represented as frequency-rank plots. Early in the evolutionary process, all networks exhibited a broad distribution of active alternatives, indicating that even the least popular options had non-trivial representation within the population. However, as the ecologies evolved, the distribution of alternatives in the spatially embedded and randomly connected populations became increasingly exponential, producing an emergent "oligopoly," in which a few entrenched local conventions eliminated all other alternatives (45, 52). Each of these conventions competed for ground against the others, but none of them assumed the majority. The ecology evolved quite differently in the homogeneously mixing populations. After the initial transient, a dominant convention rapidly emerged, breaking the symmetry with its competitors, and shifting the population into a 'winner take all' regime. Despite a large number of competing alternatives circulating in the population (see SI text), in every trial in the homogeneously mixing networks the dynamics converged on a global convention.

More generally, figure 4 shows all the replications of our study. Consistent results were found for each of the topologies at both $N$=24 and $N$=48. As a final test of increasing scale, we replicated a trial of the homogenously mixing population in which the network size was again doubled (27, 33, 53). Figure 4 shows that a shared social convention spontaneously emerged in a population of $N$=96 subjects. Global coordination in the $N$=96 population emerged on a timescale comparable to that of the initial trials ($N$=24), despite the fact that subjects had no information about how large the coordinating population was (see SI text). Within the brief timescale of the experimental observations (30 rounds of play on average), large homogenously mixing populations were significantly more likely ($p$<0.01) to spontaneously create social conventions than smaller populations with less connectivity.

**Discussion**

To ensure that our findings do not rely in any way on participants' knowledge of the size of the population or number of interaction partners, we tested the effectiveness of the informational controls used in our experimental design by providing subjects with a post-experiment survey asking them to report *i*) the most popular name in their game, *ii*) the number of people in their game, and *iii*) the number of people with whom they interacted. Across all network conditions and all network sizes, there were no significant differences in subjects' responses regarding the size of their network or the number of neighbors with whom that they interacted (see SI text). The only difference in responses was that in all the homogeneously mixing networks every respondent knew the norm, even though none of them knew how many people were using it.

We also evaluated the robustness of our results for possible biases in the initial distribution of conventions based on external focal points (54). To rule out the possibility that convergence may be biased by the pre-existing popularity of some names, we conducted controlled experiments in which participants chose their options from among an arbitrary list of ten names whose order was randomized at the beginning of the experiment (for each participant, to avoid implicit ranking effects). Results from these controlled experiments are indistinguishable from the results presented above (see Fig.4). Moreover, our findings are further supported by the observed levels of diversity in the emergent ecology of names in each of the other trials. In every trial of our study, the number of suggested names was larger than the size of the population (sometimes by more than a factor of 2; see SI text), suggesting that there were no preferred options that initially limited the set of choices in the social evolutionary process.

In sum, our findings demonstrate that social conventions can spontaneously evolve in large human populations without any institutional mechanisms to facilitate the process. Further, the results highlight the causal role played by network connectivity in the dynamics of establishing shared norms. These results contrast with prior work analyzing the effect of network structure on the speed of convergence (6, 41, 55-57). However that work focuses on the situation where there are just two competing norms that have different payoff consequences. In our case, by contrast, the number of possible norms is not fixed in advance and they all have identical payoff consequences. In this case, we find that increased network connectivity can accelerate the rate of convergence to a global norm. As a result, large populations without global information or incentives for collective agreement may nevertheless rapidly self-organize to produce universally shared collective beliefs and behaviors.

We anticipate that our results will be of interest to researchers investigating the effects of online connectedness on the emergence of new political, social and economic behaviors (58). In particular, a topic of interest for future work will be to explore the practical implications of the unintended effects of increasing social connectedness on the homogenization of behaviors and beliefs among large numbers of individuals who do not even know that they are implicitly coordinating with one another.

**Materials and Methods.**

*Experimental procedure.* Participants in the study were recruited at large from the World Wide Web. When participants arrived to play a game, they were randomly assigned to an experimental condition (i.e., a social network), and then randomly assigned to a position within that social network (see SI text). In a given round of the game, two network "neighbors" were chosen at random to play with one another. Both players simultaneously assigned names to a pictured object (e.g., a human face), blindly attempting to coordinate in the real-time exchange of naming choices (see SI text). If the players coordinated on a name, they were rewarded with a successful payment ($0.50); if they failed, they were penalized (-$0.25). (Participants could not go into debt, so failures did not incur a penalty if a participant had a balance of $0.) After a single round, the participants could see only the choices that they and their partner had made, and their cumulative pay was updated accordingly. They were then randomly assigned to play with a new "neighbor" in their social network, and a new round would begin. The object that participants were trying to name was the same for the entire duration of the game, and for all members of the game. An experimental trial concluded when all members completed the specified number of rounds. Participants did not have any information about the size the population, nor about the number of neighbors that they were connected to, nor even about which individuals they were interacting with in a given round. We explored the dynamics of convention formation over different network sizes ($24 \leq N \leq 96$) and degrees of social connectedness

(4 ≤ $Z$ ≤ $N$-1).  However, the controls within the experiment design ensured that the informational resources provided to subjects were identical across all conditions of the study.

**Acknowledgements**.

We thank H. P. Young, A. van de Rijt, L. Dall'Asta, A. Kandler, N. Perra, R. Reagans, R. Fernandez, S. Strogatz, and B. Ribeiro for helpful comments and discussion; and A. Wagner, R. Overbey, and R. Chebeleu for website development. A.B. thanks A. Vespignani and the MoBS Laboratory at Northeastern University for support in the early stages of the study, and acknowledges support from the Research Pump-Priming Fund of City University London. This work was supported in part by a James S. McDonnell Foundation grant (to D.C.).

**References**

1. Lass R (1997) *Historical linguistics and language change* (Cambridge University Press).
2. Lewis D (1969) *Convention: A philosophical study* (Blackwell).
3. Young HP (1993) The evolution of conventions. *Econometrica: Journal of the Econometric Society*:57-84.
4. Barkun M (1968) *Law without sanctions: Order in primitive societies and the world community* (Yale University Press New Haven).
5. Ehrlich PR & Levin SA (2005) The evolution of norms. *PLoS biology* 3(6):e194.
6. Young HP (1998) *Individual strategy and social structure: An evolutionary theory of institutions* (Princeton University Press).
7. Hechter M & Opp K-D (2001) *Social norms* (Russell Sage Foundation).
8. Zhang L, Zhao J, & Xu K (2014) Who create trends in online social media: The crowd or opinion leaders? *arXiv preprint arXiv:1409.0210*.
9. Boyd R (1988) *Culture and the evolutionary process* (University of Chicago Press).
10. Skyrms B (1996) *Evolution of the social contract* (Cambridge University Press).
11. Ullmann-Margalit E (1977) *The emergence of norms* (Clarendon Press Oxford).
12. Steels L & Belpaeme T (2005) Coordinating perceptually grounded categories through language: A case study for colour. *Behavioral and brain sciences* 28(4):469-488.
13. Correll SJ & Ridgeway CL (2006) Expectation states theory. *Handbook of social psychology*,  (Springer), pp 29-51.
14. Merton RK (1938) Science and the social order. *Philosophy of Science* 5(3):321-337.

15. Kearns M, Judd S, Tan J, & Wortman J (2009) Behavioral experiments on biased voting in networks. *Proceedings of the National Academy of Sciences* 106(5):1347-1352.
16. Young MW (1971) Fighting with food. Leadership, values and social control in a Massim society. *Fighting with food. Leadership, values and social control in a Massim society.*
17. Salganik MJ, Dodds PS, & Watts DJ (2006) Experimental study of inequality and unpredictability in an artificial cultural market. *Science* 311(5762):854-856.
18. Bicchieri C (2006) *The grammar of society: The nature and dynamics of social norms* (Cambridge University Press).
19. Hayek FA (1960) *The constitution of liberty* (Routledge).
20. Sugden R (1989) Spontaneous order. *The Journal of Economic Perspectives*:85-97.
21. Haken H (1988) *Information and self-organization: A macroscopic approach to complex systems* (Springer).
22. Strogatz S (2003) *Sync: The emerging science of spontaneous order* (Hyperion).
23. Steels L (1995) A self-organizing spatial vocabulary. *Artificial life* 2(3):319-332.
24. Judd S, Kearns M, & Vorobeychik Y (2010) Behavioral dynamics and influence in networked coloring and consensus. *Proceedings of the National Academy of Sciences* 107(34):14978-14982.
25. Young HP (1998) Social norms and economic welfare. *European Economic Review* 42(3):821-830.
26. Helbing D, Yu W, Opp K-D, & Rauhut H (2014) Conditions for the Emergence of Shared Norms in Populations with Incompatible Preferences. *PloS one* 9(8):e104207.
27. Baronchelli A, Felici M, Loreto V, Caglioti E, & Steels L (2006) Sharp transition towards shared vocabularies in multi-agent systems. *Journal of Statistical Mechanics: Theory and Experiment* 2006(06):P06014.
28. Durrett R & Levin SA (2005) Can stable social groups be maintained by homophilous imitation alone? *Journal of Economic Behavior & Organization* 57(3):267-286.
29. Galantucci B (2005) An experimental study of the emergence of human communication systems. *Cognitive science* 29(5):737-767.
30. Selten R & Warglien M (2007) The emergence of simple languages in an experimental coordination game. *Proceedings of the National Academy of Sciences* 104(18):7361-7366.
31. Garrod S & Doherty G (1994) Conversation, co-ordination and convention: An empirical investigation of how groups establish linguistic conventions. *Cognition* 53(3):181-215.
32. Anderson PW (1972) More is different. *science* 177(4047):393-396.
33. Castellano C, Fortunato S, & Loreto V (2009) Statistical physics of social dynamics. *Reviews of modern physics* 81(2):591.
34. Lehmann J, Gonçalves B, Ramasco JJ, & Cattuto C (2012) Dynamical classes of collective attention in twitter. *Proceedings of the 21st international conference on World Wide Web*, (ACM), pp 251-260.

35. Kooti F, Yang H, Cha M, Gummadi PK, & Mason WA (2012) The Emergence of Conventions in Online Social Networks. *in Proceedings of the Sixth International AAAI Conference on Weblogs and Social Media (ICWSM)*.
36. Lin Y-R, Margolin D, Keegan B, Baronchelli A, & Lazer D (2013) #Bigbirds Never Die: Understanding Social Dynamics of Emergent Hashtags. *in Proceedings of the Seventh International AAAI Conference on Weblogs and Social Media (ICWSM)*:370-379.
37. Helbing D & Yu W (2009) The outbreak of cooperation among success-driven individuals under noisy conditions. *Proceedings of the National Academy of Sciences* 106(10):3680-3685.
38. Hume D (1748) *A treatise of human nature* (2012. Courier Dover Publications).
39. Wittgenstein L (1958) *Philosophical investigations* (Blackwell Oxford).
40. Blume LE (1993) The statistical mechanics of strategic interaction. *Games and economic behavior* 5(3):387-424.
41. Kreindler GE & Young HP (2014) Rapid innovation diffusion in social networks. *Proceedings of the National Academy of Sciences* 111(Supplement 3):10881-10888.
42. Dall'Asta L, Baronchelli A, Barrat A, & Loreto V (2006) Nonequilibrium dynamics of language games on complex networks. *Physical Review E* 74(3):036105.
43. Klemm K, Eguíluz VM, Toral R, & San Miguel M (2003) Nonequilibrium transitions in complex networks: A model of social interaction. *Physical Review E* 67(2):026120.
44. Centola D, Gonzalez-Avella JC, Eguiluz VM, & San Miguel M (2007) Homophily, cultural drift, and the co-evolution of cultural groups. *Journal of Conflict Resolution* 51(6):905-929.
45. Baronchelli A, Dall'Asta L, Barrat A, & Loreto V (2006) Topology-induced coarsening in language games. *Physical Review E* 73(1):015102.
46. Centola D (2010) The spread of behavior in an online social network experiment. *science* 329(5996):1194-1197.
47. Centola D (2011) An experimental study of homophily in the adoption of health behavior. *science* 334(6060):1269-1272.
48. Albert R & Barabási A-L (2002) Statistical mechanics of complex networks. *Reviews of modern physics* 74(1):47.
49. Barth F (1970) *Ethnic groups and boundaries* (Universitets Forlaget).
50. Axelrod R (1986) An evolutionary approach to norms. *American political science review* 80(04):1095-1111.
51. Cassidy FG (1985) *Dictionary of American Regional English. Volume I: Introduction and AC* (Belknap, Cambridge, MA).
52. Ericson R & Pakes A (1995) Markov-perfect industry dynamics: A framework for empirical work. *The Review of Economic Studies* 62(1):53-82.
53. Jackson MO (2010) An overview of social networks and economic applications. *The handbook of social economics*:511-585.
54. Schelling TC (1978) Altruism, meanness, and other potentially strategic behaviors. *American Economic Review* 68(2):229-230.
55. Ellison G (1993) Learning, local interaction, and coordination. *Econometrica: Journal of the Econometric Society*:1047-1071.
56. Montanari A & Saberi A (2010) The spread of innovations in social networks. *Proceedings of the National Academy of Sciences* 107(47):20196-20201.

57. Young HP (2011) The dynamics of social innovation. *Proceedings of the National Academy of Sciences* 108(Supplement 4):21285-21291.
58. Backstrom L, Huttenlocher D, Kleinberg J, & Lan X (2006) Group formation in large social networks: membership, growth, and evolution. *in Proceedings of the 12th ACM SIGKDD international conference on Knowledge discovery and data mining*:44-54.

**Figures**

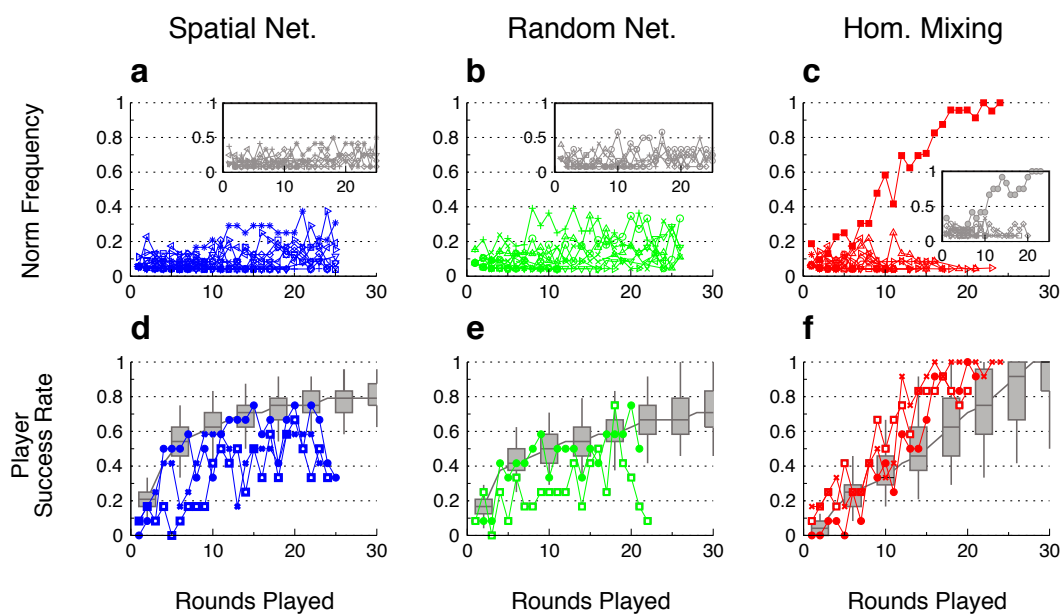

**Figure 1. The evolution of social conventions across network topologies ($N$=24).** Panels indicate spatially embedded networks (**a**, **d - blue**), random networks (**b**, **e - green**), and homogeneously mixing populations (**c**, **f - red**), for eight independent trials of the study ($N$=24). Panels **a, b** and **c** show the evolving ecology of norms for representative trials from each condition. Insets show representative model simulations. The corresponding time series (**d**, **e**, **f**) show the average level of successful matching among individual players. Model results are shown in grey (95% confidence intervals over 10,000 realizations). In spatially embedded networks (**a**, **d**), players achieved moderate success with local conventions, creating regional competition, and preventing a single convention from emerging across the population. Similarly, in random networks (**b**, **e**) moderately successful local coordination produced groups in the networks, but no global consensus. By contrast, in homogeneously mixing populations (**c**, **f**), initial local failures resulted in rapid population-level learning and global coordination on a single convention.

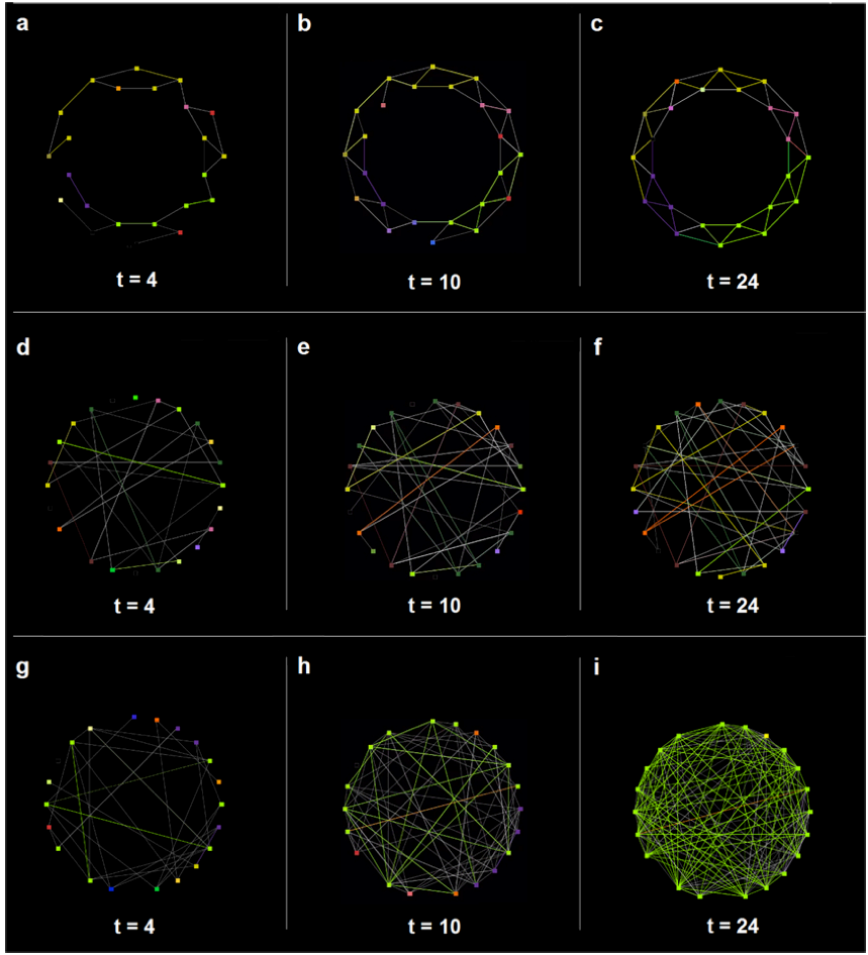

**Figure 2. Temporal dynamics of convention formation in representative experimental trials.** Panels correspond to a spatially embedded network (**a-c**), a random network (**d-f**), and a homogeneously mixing population (**g-i**) (*N*=24). Each color corresponds to a unique name used in the trial. Node color refers to the name that was most recently used by that actor, and edge color indicates the name that the two players matched on in their most recent interaction. A white edge indicates that the two players failed to match in their most recent interaction. In the spatial network (**a-c**), local interactions produce clusters of coherent coordination around a shared convention, with contested border regions. A similar dynamic unfolds in the random network (**d-f**), where repeated interaction leads to local coordination. In the homogeneously mixing population (**g-i**), a single name assumes dominance, becoming the global convention.

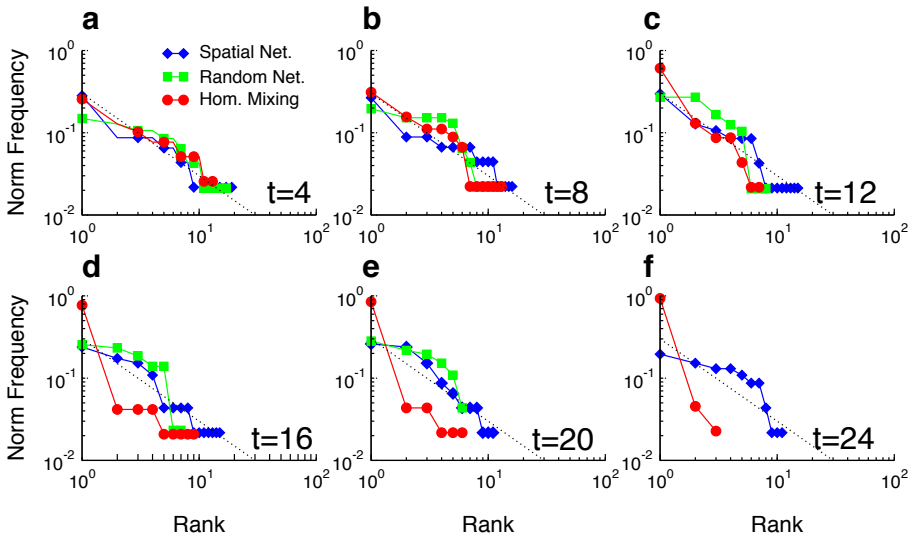

**Figure 3. The shifting distributions of conventions in diverse network topologies.** Frequency-rank plots for spatially embedded networks (blue), random networks (green), and homogeneously mixing populations (red) ($N=48$, single trial from each condition). In the initial rounds of the game (**a**, **b**, **c**), the fraction of the population using each name scales as a broad distribution with the overall rank of the name. The scaling is similar in all networks. However, (**d**) by round 16 in the homogeneously mixing population a single name breaks the symmetry with the others, accelerating its growth and driving the other options toward extinction (**e**, **f**). Over the same time interval, the spatial network and random network exhibit much more moderate growth dynamics, producing exponential distributions with a few competing dominant groups, but without any "winner" emerging. The dashed line in all panels provides a visual guide for the Zipf distribution (slope -1). Note that for t=24 the curve for random networks is absent due to a shorter experimental time.

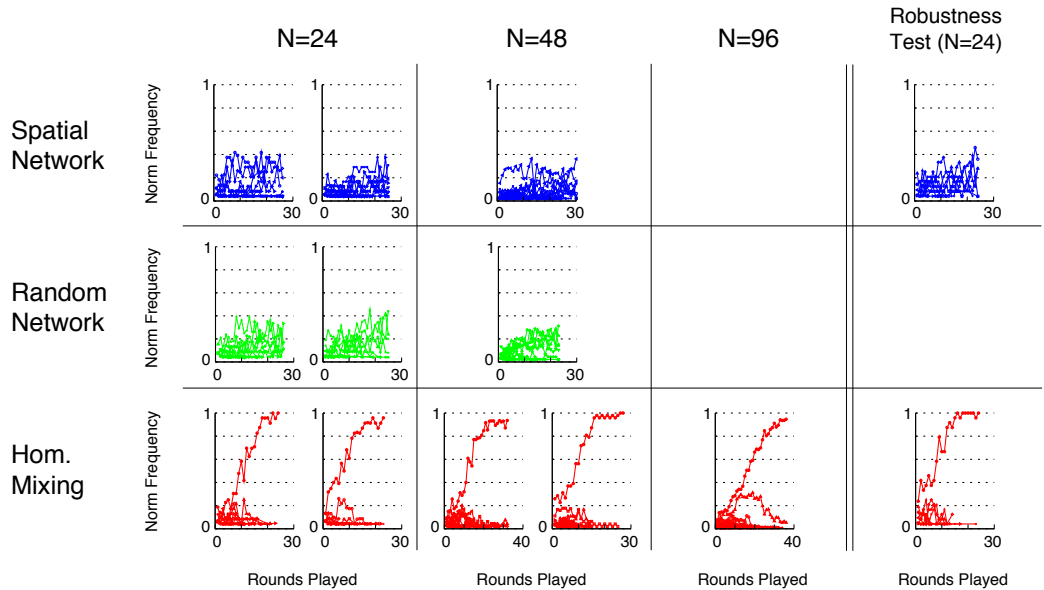

**Figure 4. Size of dominant convention across conditions.** The temporal evolution of the ecology of norms is reported for 13 experimental trials. Results show a significant difference in dominant convention size between locally connected networks (spatial and random networks, combined) and homogenously mixing populations (Wilcoxon Rank Sum Test, $p<0.01$). In the spatial network and random network trials, the most popular convention was never used by more than 45% of the group, and in most trials was well below 40%. Homogenously mixing populations, on the other hand, produced global social conventions that reached up to 100% of the population. More generally, in spatial networks the average size of the dominant convention across trials was 30% of the population (including $N$=24 and $N$=48), in random networks the average size was 33% (including $N$=24 and $N$=48), and in homogenously mixing populations, the average size was 96% (including $N$=24, $N$=48 and $N$=96).

# Supplementary Information

(Please note a new reference numbering for the SI)

**Experimental Design.**

Each trial of the study consisted of a network structure, a pre-specified number of rounds, and a set of participants equal to the size of the network (*N*). Participants were randomly assigned to a trial (i.e., an active network), and then they were randomly assigned to a node within that network (see Figure S1). Participants had no information about who, or how many individuals they were directly connected to, or how many people were in the population. The subject experience was identical in every network condition. Consequently, any differences in collective behaviour across network conditions were due to the structure of the interaction networks, and not to information the subjects had about the structure or size of the population that they were in.

**Subject Experience.**

Each round, subjects were randomly paired with one of their network neighbors, and shown a picture for which they had to enter a name (Figure S2.a). They were given a 20 second time limit in which they could enter a name for the pictured face. During the same time interval, their partner was given the same face, and the same time limit. If either subject did not complete an answer in the allotted time, the system registered a void answer, and that round was considered a "null" round in which no information was exchanged. In terms of the participants' scores, both participants were registered as a failed interaction for a given round if one or both of them produced a void answer. Alternatively, if both players entered a name for the pictured face within the time limit, the round concluded with a page that told both subjects the name their partner entered, and indicated whether they successfully matched or not. Their score for that round was indicated as either a "match" or a "no match," respectively, depending on whether they were successful or not. Accordingly, the players' scores were also updated based on whether they succeeded or not (Figure S2.b). The players then waited to be assigned with a new partner (another one of their network neighbors). Once a player was assigned, the player was again presented with same game screen and the opportunity to name the image, within the 20 second time interval (Figure S2.c). This procedure was repeated for the allotted number of rounds until the player had completed the pre-specified number of rounds.

Each player was permitted to move at their own pace through the game. Some players may therefore have completed their allotted number of rounds before some of their network neighbors. To ensure that none of these neighbors was "stranded" with all of their neighbors finishing before they could complete the pre-specified number of rounds, some players were given the opportunity to play additional rounds of the game until the slower player had completed her full allotment of rounds. These additional rounds were identical to the earlier rounds, and ensured that by the end of the game, every subject had played at least the pre- specified number of rounds. When all subjects had completed the pre-specified number of rounds, the game ended.

**Subject Recruitment.**

Participants in our study were recruited at large from the World Wide Web to be players in the "Name Game". When they arrived to the study, each participant completed the registration by choosing a username and an avatar. Participants were then provided with a specific time to return to the site when they would play the live Name Game with a group of anonymous, randomly selected subjects. The study was run for a 140-day period, over which time recruitment campaigns were conducted to attract subjects to participate in the study. In total, 510 subjects were recruited to participate in the study. 120 of them participated in spatially embedded lattice networks, 96 of them participated in randomly connected graphs, and 264 of them participated in homogeneously mixing networks, as detailed below in the Data Analysis and Replication sections. Subjects were recruited through email advertisements sent to a broad list of websites that subscribe to the Adweek mailing list.

**Network Structures.** Each trial of the study consisted of a network structure and a population that filled it. We explored three different topologies within our design. 1) A spatially embedded network, which was structured as a one-dimensional lattice with degree 4. Each node was tied to nearest neighbors and next-nearest neighbors. 2) A random network topology, which was structured as a random graph with constant degree 4. 3) A homogeneously mixing population, or a complete graph. Each of these topologies was tested for populations of size $N$=24 and $N$=48. For $N$=96, we conducted a single study with the homogeneously mixing network (See Replication section). As discussed in the Model section below, our formal predictions for $N$=96 were qualitatively the same those found for $N$=48 and $N$=24.

**Data Analysis.**

The experimental data were produced as a chronological sequence of rounds, ordered according to the starting time of the game (Figure S3). Each round consisted of two participants each typing a name. As described above, users had 20 seconds to type a name, after which time the system registered their answer as void. A round was considered successful only when the two participants entered the same name, irrespective of case. In the analysis, the population's evolutionary time scale is measured in terms of the number of times the entire population goes through a single round of the game, which we refer to as a single "Round Played." A Round Played for the population corresponds to $N/2$ individual rounds in the data sequence. Thus, a Round Played is equivalent to $N/2$ pairs (i.e., the entire population) all playing once. This approach allows us to measure the movement of the entire population through sequential rounds of the Name Game, based on the standard mapping between Monte Carlo steps and microscopic interactions (1). The value of each measured quantity for the population at any Round Played is obtained by averaging over the last $N/2$ individual rounds. Thus, for example, the average success rate of the population at the *(x+1)*th Round Played corresponds to the average of the individual outcomes in the interval between the lines *(xN/2+1)* and *(x+1)N/2* in the data sequence (success being a binary variable, is assigned values 0 or 1 at each individual round) (Figure S3). Analogously, the frequency of each norm at the *(x+1)*th Round Played is relative to the frequency of the other norms in the same interval. If either member of a given pair fails to type a name within an interaction, then this interaction is treated as void.

*Treatment of void answer field.* If either member of a given pair produces an empty field within an interaction, then this interaction is treated as void. This is handled differently for the analysis of success versus the analysis of norm ecology. For success rates, a void entry from either member of a pair indicates non-participation (i.e., a null interaction) in that round. Thus, in the evaluation of the temporal evolution of the success rate we disregarded all those interactions in which void entries appears. For analyzing the evolving norm ecology, however, if only one member of a pair produced a void field while the other member typed a name, we include the name that was typed by the active participant as a data point in our analysis of the active names within the population. As a consequence, for a single experimental realization, the considered temporal sequence can be slightly different for the series of pairwise success and that of the overall norm ecology. It is worth stressing that all results are robust against variations of the specific procedure adopted to take into account the void field, and that different ways of dealing with interactions in which one of the participants did not type any guess produce results that are equivalent, and virtually undistinguishable, under any respect.

**Replication.**

The results presented in the main text were consistent across all replications of our experiments. We replicated the experiment 8 times for homogeneous mixing populations ($N$=12, 24, 48, 96), 6 times for the spatial networks ($N$=12, 24, 48) and 3 times for the random networks ($N$=24, 48). All trials of size $N$=24 were run for an average of 25 rounds; trials of $N$=48 were run for an average of 30 rounds, and the trial of $N$=96 was run for 40 rounds.

The choice of the above-mentioned trials was dictated by our research questions. As predicted by the model, small $N$ trials ($N$=12) produced similar dynamics across all experimental conditions, preventing any identification of the effects of network structure on convention formation. Six trials were conducted with $N$=24. According to our model, this was the minimal population size at which significant differences in the emergent dynamics of local coarsening versus symmetry breaking could be detected. For robustness, each trial was replicated twice in each condition. This allowed identification of the main dynamical differences across conditions and assessment of the validity of the model predictions, as shown in Figure 1 of the main text. The replications with $N$=48 corroborated these results in larger networks and allowed us to obtain data sufficient to construct the distributional analyses presented in Figure 3 of the main text.

Because consensus becomes more difficult with increasing scale, the failures in the spatial lattices and random networks indicated that larger N studies with those topologies would yield similar results. Our focus for additional replications was thus the homogeneously mixing population, i.e., the only condition in which a global consensus emerges in our experiments. We replicated the $N$=48 homogenously mixing experiment a second time, and additionally we tested our results in a final, considerably more demanding trial with 96 participants. As a final test of our findings, we also conducted two "Name List" trials, which served as a further check on the robustness of the results (see "Robustness" section below).

**Robustness**.

A possible concern with the design of our study is that the distribution of words entered by subjects would be skewed in favor of a particularly salient name (where saliency could have been due to a vast range of external events/factors, such as the celebrities in the news, etc.), which would drive convergence by artificially reducing the number of options in the population. To check the robustness of our results in a setting that eliminated these concerns, we replicated our study in homogeneous mixing and spatial networks of size $N$=24, in an environment in which participants could not type their own name entries. Instead of allowing participants to enter their proposed name in a text box, we provided them a fixed list of 10 names. Participants had to name a feminine face, and could select in each round one name from the fixed list of: Sophia, Emma, Isabella, Olivia, Ava, Emily, Abigail, Mia, Madison, and Elizabeth; corresponding to the most popular baby names for females for 2012 in the U.S. according to the U.S. Social Security Office (2). The order of these names was randomized at the beginning of the experiment for each participant, in order to rule out possible ordering biases. Figure S4 (bottom row) shows that the results are consistent with those obtained in the trials using open field name entry.

As a secondary check on our results, we examined the data to determine whether the list of actual names that were suggested by subjects in any of the trials was artificially limited to a small number of options. As shown in Figure S4**,** in each of the conditions with open name fields, the number of names entered by subjects was greater than the size of the population in the game.

**Post-trial user tests.**

To ensure that the informational controls in our study were effective, we provided subjects from five selected trials with a post-study questionnaire, asking them to report 1) the number of people in their game, and 2) the number of people with whom they directly interacted. Figure S5(a) shows the mean and standard deviation of responses to the number of people in the game (normalized by the number of rounds that subjects played). There were no significant differences in the average responses across network structures and network sizes ($p > 0.2$, Mann-Whitney U-Test). Similarly, Figure S5(b) shows the normalized mean and standard deviation of responses for the number of people that players believed they interacted with. There were no significant differences in the average responses across network structures and network sizes ($p > 0.2$, Mann-Whitney U-Test).

**Limitations of the Study.**

As with all experiments, the scientific controls that made this study possible also present limitations. Most notably, practical constraints on the number of people that can be recruited to simultaneously participate in an evolving social convention within an experimental environment prevented us from running larger experiments (i.e., $N > 100$). These constraints also limited the duration of our experiments – i.e., the number of rounds of play – since our design relied on subjects' sustained behavioral engagement over the entire study. While these practical constraints limited the size of our empirical study, the correspondence between our model and the experimental data provides guidance for our expectations about how these evolutionary systems behave at larger sizes and longer time scales. As discussed below in the Model section, the results from our simulations suggest convergence time in each of the three networks scales as a direct function of the topology. For spatial lattices, convergence is expected to scale as $O(N^2)$ Rounds Played, while for both the random graph and the homogeneously mixing population, convergence time is expected to scale as $O(N^{0.5})$ Rounds Played. Based on the agreement between our experimental results and the model, we speculate that the dynamics of norm evolution within each network topology will follow the patterns of coarsening (in the spatial lattice and early stages of the random graph) and symmetry breaking (in the homogeneously mixing population and late stages of the random graph), as described below.

**Model.**

The Naming Game model constructs a population of $N$ agents that engage in pairwise interactions in order to negotiate local coordination, and is able to demonstrate the emergence of a global convention among them (3). An example of such a game is that of a population that has to reach consensus on the name for an object, using only local interactions, as in the Name Game experiment.

In the model, each agent has an internal name inventory in which an *a priori* unlimited number of words can be stored. As an initial condition, all inventories are empty. At each time step, a pair of agents is chosen randomly, one playing as "speaker", the other as "hearer", and interact according to the following rules:

- The speaker randomly selects one of her words (or invents a new word if her inventory is empty) and conveys it to the hearer;
- If the hearer's inventory contains such a word, the two agents update their inventories so as to keep only the word involved in the interaction (*success*);
- Otherwise, the hearer adds the word to those already stored in her inventory (*failure*).

The non-equilibrium dynamics of the Naming Game are characterized by three temporal regions: (i) initially the words are invented; (ii) then they spread throughout the system inducing a reorganization process of the inventories; (iii) this process eventually triggers the final convergence towards the global consensus (all agents possess the same unique word). The dynamics leading to final consensus, and the associated scaling of the consensus time with the population size, depend crucially on the topological properties of the social network identifying the set of possible interactions among individuals.

*Homogeneous mixing populations*

In homogeneously mixing populations, step (iii) above is triggered by a symmetry breaking in the ecology of conventions, in which the most popular norm will progressively eliminate all the competitors. For a population of size *N*, consensus is reached in a time $t_{conv} \sim O(N^{1.5})$ microscopic interactions, i.e. in $O(N^{0.5})$ Rounds Played, according to our definition (3).

The symmetry breaking process has been clarified analytically by considering a system prepared in an initial configuration in which half of the population knows only convention "A" and the other half only convention "B" (3, 4). Here, stochastic fluctuations break the initial symmetry between the two norms making one of them more popular, and the interaction dynamics amplify this small advantage until a final state in which the initially disadvantaged convention is extinct. Thus, in the limit of large population, any initial imbalance in favour of one of the two conventions will eventually determine the success of that convention (4, 5).

The situation is more complex when more than two conventions are present in the system, but the overall symmetry-breaking picture remains the same (3, 5). This mechanism is radically different from what is observed in pure imitation models, such as the Moran process or the voter model, where fluctuations dominate the whole of the process leading to consensus, and the advantage of one of the competing states can be reversed easily during the dynamics of the process (6). When only pure imitation is at work, the fluctuation-driven consensus is reached in $O(N)$ Rounds Played (or Monte-Carlo steps).

*Networked populations*

In the model described above, at each time step two agents are randomly selected. The assumption behind this homogeneous mixing, or "mean-field", rule is that the population is not structured and that any agent can in principle interact with any other. However, when actors are embedded in a fixed network, the topology in which the population is embedded identifies the set of possible interactions among the individuals. Thus, the group of communicating individuals can be described as a network in which each node represents an agent and the links connecting different nodes determine the allowed communication channels. The (statistical) properties of the underlying network significantly affect the overall dynamics of the model.

*Lattices.* On low-dimensional lattices each agent can rapidly interact two or more times with its neighbors, favoring the establishment of a local consensus with a high success rate, i.e. of small sets of neighboring agents sharing a common unique word. As the process evolves, these "clusters" of neighboring agents with a common unique word undergo a coarsening phenomenon with a competition among them driven by the fluctuations of the interfaces. The coarsening picture can be extended to higher dimensions, and the scaling of the convergence time has been shown to be $O(N^{2/d})$, where $d \leq 4$ is the dimensionality of the space (7). This prediction has been confirmed numerically.

*Small-world networks.* Results concerning the homogeneously mixed population, on the one hand, and regular lattices, on the other, act as fundamental references to understand the role of the different properties of complex networks. In between these regimes, the small-world network (8) allows us to interpolate progressively from regular structures to random graphs by tuning the *p* parameter describing the probability that a link of the regular structure is rewired to a random destination. The main result is that the presence of shortcuts, linking agents otherwise far from each other, allows recovering the fast convergence typical of the mean-field case. The finite connectivity, on the other hand, guarantees that there will be a good degree of coordination between neighbors from the start of the dynamics, as in regular structures.

In these randomized topologies, two different regimes are observed (9). For times shorter than a cross-over time, $t_{cross} = O(N/p^2)$, one observes the usual coarsening phenomena since the clusters are typically one-dimensional, i.e., since the typical cluster size is smaller than 1/p. For times much larger than $t_{cross}$, the dynamics shift. They become dominated by the existence of shortcuts, and follow the mean-field behavior similar to the one observed on the complete graph. The convergence time, measured in microscopic interactions, scales therefore as $N^{1/2}$ and not as $N^{2/d}$ (as in low-dimensional lattices) (9).

*Complex networks.* Most of the relevant features exhibited by complex networks have been explored systematically, mainly by means of computer simulations. The scaling exponents observed in both homogeneous (eg Erdös-Rényi (10)) and heterogeneous (eg Barabási-Albert (11)) networks are similar to the one observed in the Watts-Strogatz small-world graphs (8) for both consensus time and memory usage. In particular, the scaling laws observed for the convergence time is a general robust feature that is not affected by further topological details, such as the average degree, the clustering or the particular form of the degree distribution (12).

*3. Robustness of the model*

The model described above has been modified in several directions to test its robustness (6, 13-21). For example, the rule describing how a word is selected from the inventory has been investigated, and more efficient strategies have been identified (13). In the same way, the symmetric update of the inventories has been altered and the role of the feedback between the agents has been investigated (16, 21). However, all permutations of the model exhibit qualitatively similar dynamics, which rely on two essential elements: 1) memory and 2) the fact that for a success to take place both agents must have already heard the successful convention in the past. That is, actors depend upon multiple exposures to a term in order to successfully coordinate on it (22). These two elements seem to be crucial to reproduce the observed dynamics of coarsening on low dimensional lattices and symmetry breaking in random networks and homogeneously mixing populations.

**Model rules and user behavior.**

To test the relationship between the model and the experimental results in more depth, we investigated how well the individual behavior of participants in the study matched with the theoretical model. We then simulated the long term dynamics of the real user behaviors, and compared it to the expected dynamics from the theoretical model.

In the theoretical model, agents accrue a list of word options based on their history of interactions. The only words that they can enter at a given round are those that currently exist in their inventory. If they experience a successful match, their inventory is deleted except for the matching word. The inventory can increase again through subsequent interactions, however any subsequent matches again reset the inventory to 1, leaving only the most recent matching word. We evaluated subjects' behavior in terms of whether the answers they provided at every round were consistent with answers that agents in our theoretical model could have provided; i.e., we evaluated whether the answers that subjects actually used would have been in their "inventories" if they had followed the same rules as the model, given their histories of past interactions, failures and successes.

Remarkably, we found a 95% agreement between the model and the subjects' behaviors. In other words, 95% of the time, subjects' choices were entirely consistent with the rules of the theoretical model. When these individual dynamics were simulated (95% model rules, 5% random entries – either through novel word choice, or through choosing words from a deleted inventory), the collective dynamics were indistinguishable from those of the theoretical model. Consistent with the model, these results suggest that subjects' behaviors were governed more by their recent successes than by their history of past plays.

**Model implications.**

As shown in the main text, the model captures well the results of the Name Game experiment, and the microscopic rules provide a good fit with the empirically measured user behavior. However, as discussed above, experimental constraints limit the region of accessible parameters, in particular with respect to the duration of an experiment and the population size. The model behavior allows us to make grounded predictions on the outcome of experiments at larger scales. In particular we would expect that:

a) The dynamics observed on the random graph would eventually be different from the one of spatial networks, and the scaling of the convergence time with the population size will be similar to the one observed in the homogeneously mixing population (9).

b) The difference between the initial stages of the spatial graph and the homogeneously mixing population will be more and more significant (12).

c) The symmetry breaking transition, which governs the consensus process in the homogeneously mixing population will result in a characteristic S-shaped behavior of the success rate curve (3).

Figure S6 shows the evolution of the space of norms for the three networks considered in the main text, (a,d) spatial lattice networks, (b,e) random graphs, and (c,f) homogeneously mixing, for populations of $N$=48. Panels a-c show the results for the experimentally accessible regions of the dynamics (i.e., 30 Rounds). Panels d-f show the same simulations with time scales extended until final convergence. To demonstrate these effects at larger population scales, Figure S7 shows the evolution of the player success rate until model convergence in all three network conditions for populations of size of $N$=1000.

**SI References**

1. Marro J & Dickman R (1999) *Nonequilibrium phase transitions in lattice models* (Cambridge University Press).
2. 2013) Top 10 Baby Names For 2012. *Official Social Security Website (*http://www.ssa.gov/OACT/babynames/).
3. Baronchelli A, Felici M, Loreto V, Caglioti E, & Steels L (2006) Sharp transition towards shared vocabularies in multi-agent systems. *Journal of Statistical Mechanics: Theory and Experiment* 2006(06):P06014.
4. Baronchelli A, Loreto V, & Steels L (2008) In-depth analysis of the Naming Game dynamics: the homogeneous mixing case. *International Journal of Modern Physics C* 19(05):785-812.
5. Baronchelli A, Dall'Asta L, Barrat A, & Loreto V (2007) Nonequilibrium phase transition in negotiation dynamics. *Physical Review E* 76(5):051102.
6. Castellano C, Fortunato S, & Loreto V (2009) Statistical physics of social dynamics. *Reviews of modern physics* 81(2):591.
7. Baronchelli A, Dall'Asta L, Barrat A, & Loreto V (2006) Topology-induced coarsening in language games. *Physical Review E* 73(1):015102.
8. Watts DJ & Strogatz SH (1998) Collective dynamics of 'small-world' networks. *nature* 393(6684):440-442.
9. Dall'Asta L, Baronchelli A, Barrat A, & Loreto V (2006) Agreement dynamics on small-world networks. *EPL (Europhysics Letters)* 73(6):969.
10. Erdos P & Rényi A (1959) On random graphs. *Publicationes Mathematicae Debrecen* 6:290-297.
11. Barabási A-L & Albert R (1999) Emergence of scaling in random networks. *science* 286(5439):509-512.
12. Dall'Asta L, Baronchelli A, Barrat A, & Loreto V (2006) Nonequilibrium dynamics of language games on complex networks. *Physical Review E* 74(3):036105.
13. Baronchelli A, Dall'Asta L, Barrat A, & Loreto V (2005) Strategies for fast convergence in semiotic dynamics. *Artificial Life X - Proceedings of the Tenth International Conference on the Simulation and Synthesis of Living Systems*:480-485.
14. Wang W, Lin B, Tang C, & Chen G (2007) Agreement dynamics of finite-memory language games on networks. *The European Physical Journal B* 60(4):529-536.

15. Dall'Asta L & Castellano C (2007) Effective surface-tension in the noise-reduced voter model. *EPL (Europhysics Letters)* 77(6):60005.
16. Lu Q, Korniss G, & Szymanski BK (2008) Naming games in two-dimensional and small-world-connected random geometric networks. *Physical Review E* 77(1):016111.
17. Brigatti E (2008) Consequence of reputation in an open-ended naming game. *Physical Review E* 78(4):046108.
18. Brigatti E & Roditi I (2009) Conventions spreading in open-ended systems. *New Journal of Physics* 11(2):023018.
19. Lipowski A & Lipowska D (2009) Language structure in the n-object naming game. *Physical Review E* 80(5):056107.
20. Lei C, Jia J, Wu T, & Wang L (2010) Coevolution with weights of names in structured language games. *Physica A: Statistical Mechanics and its Applications* 389(24):5628-5634.
21. Baronchelli A (2011) Role of feedback and broadcasting in the naming game. *Physical Review E* 83(4):046103.
22. Centola D & Macy M (2007) Complex contagions and the weakness of long ties1. *American Journal of Sociology* 113(3):702-734.

**SI text Figures**

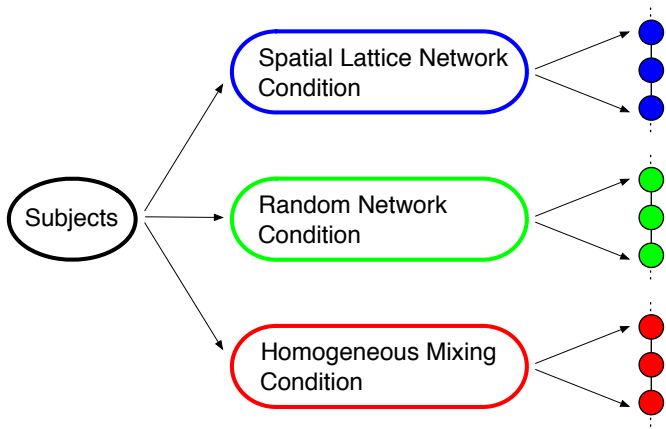

**Figure S1. Schematic representation of randomization to conditions.** Subjects arriving to the study were first randomly assigned to an experimental condition (i.e., a social network), and then randomly assigned to a specific node within that network. The nodes directly connected to an individual constituted the set of potential partners that she could interact with during the game.

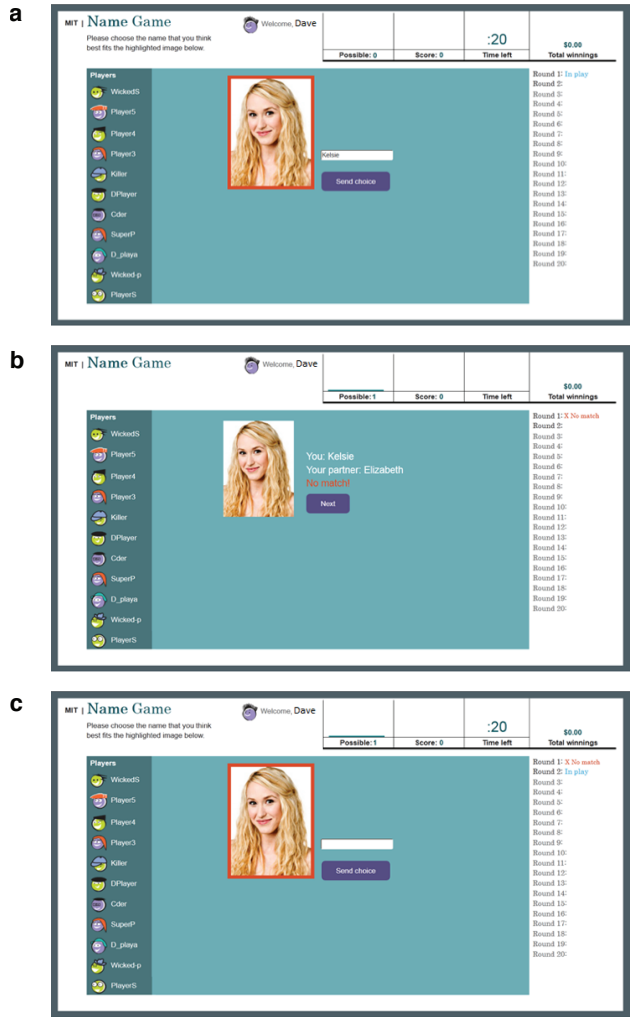

**Figure S2. User interface and experience.** At the start of the study, **(a)** subjects are given a picture of a face and an entry field (with no character limit) to provide a name. Once a subject submits her choice **(b)** she is then exposed to the choice of her interaction partner. If the choices match, both subjects are rewarded, otherwise there is no reward for that round. The study then advances to the next round **(c)**, where the player is again matched with a partner and repeats the same procedure. Each round lasts for 20 seconds maximum, and the players have real-time information on their record of matches and failures over the past rounds. The 'Players' column on the left of the screen is a static representation of other player icons, identical across experimental conditions (and hence independent from the actual topology and population size).

| Round | Player 1 | Name 1 | Player 2 | Name 2 | Success | Round Played | Player Success Rate |
|---|---|---|---|---|---|---|---|
| 1 | A | "Sarah" | L | "Isabella" | 0 | | |
| 2 | F | "Maria" | B | "Anna" | 0 | | |
| 3 | G | "Isabella" | I | "Isabella" | 1 | | |
| 4 | D | "Lauren" | C | "Sarah" | 0 | | |
| 5 | E | "Giulia" | A | "Anne" | 0 | 1 | 1/5 |
| 6 | F | "Mary" | H | "Mary" | 1 | | |
| ... | ... | ... | ... | ... | ... | | |

**Figure S3. Schematic representation of the data for a population of size N=10.** Every experiment generates an ordered list of individual rounds, ordered on the basis of their starting time. If the names typed by the two players are the same, the interaction is a success, and the success variable takes value 1, otherwise it is a failure and the relative variable is set at 0. Global quantities, such as the Player Success Rate in the figure, are averaged over N/2 individual rounds, corresponding to one Round Played. In the figure, the Player Success Rate in the first Round Played success is equal to 1/5 since one pair out of five achieved success. Within each Round Played, each player plays once on average. For instance, in Round Played 1 user A plays twice (rounds 1 and 5) and player H does not play at all, while in Round Played 2, it would be reversed.

|  | Cumulative Number of Alternatives Created In Trial |
|---|---|
| Trial 1  (N=24, Spatial Network) | 53 |
| Trial 2  (N=24, Spatial Network) | 83 |
| Trial 3  (N=24, Random Network) | 40 |
| Trial 4  (N=24, Random Network) | 30 |
| Trial 5  (N=24, Homophilous Mixing) | 50 |
| Trial 6  (N=24, Homophilous Mixing) | 48 |
| Mean | 50.66 |
| Standard Deviation | 17.88 |
| Trial 7  (N=48, Spatial Network) | 66 |
| Trial 8  (N=48, Random Network) | 51 |
| Trial 9  (N=48, Homophilous Mixing) | 64 |
| Trial 10  (N=48, Homophilous Mixing) | 52 |
| Mean | 58.25 |
| Standard Deviation | 7.84 |
| Trial 11  (N=96, Homophilous Mixing) | 120 |
| Mean | - |
| Standard Deviation | - |

**Figure S4. Cumulative names entered over the course of the study.** The numbers reported here indicate the number of different words in active circulation each population. Identical spellings with different cases were considered to be the same word. The number of alternative names created in each trial was larger than the population size.

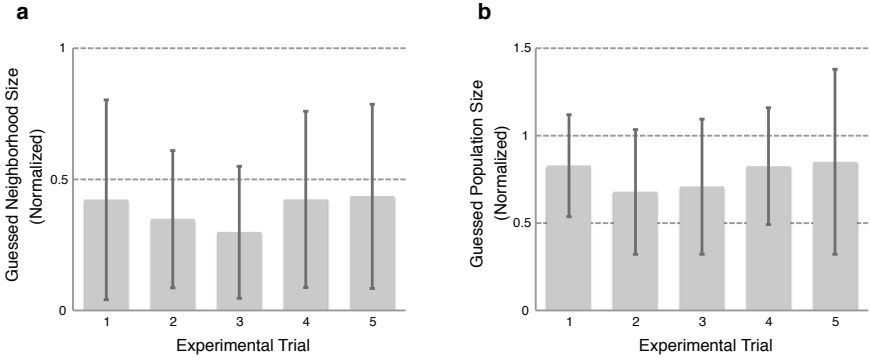

**Figure S5. Subject's informational awareness about experimental conditions. (a)** Subjects' reported beliefs about the number of people they interacted with are shown for five experimental trials (normalized by the number of rounds that subjects played). Results are shown for a representative (1) Spatial Network $N=24$, (2) Homogeneous Mixing $N=24$, (3) Random Network $N=48$, (4) Homogeneous Mixing $N=48$, and (5) Homogeneous Mixing $N=96$. For the same trials **(b)** shows subjects' reported beliefs about the population size of their study (normalized by the number of rounds that subjects played). Error bars indicate the standard deviation in responses.

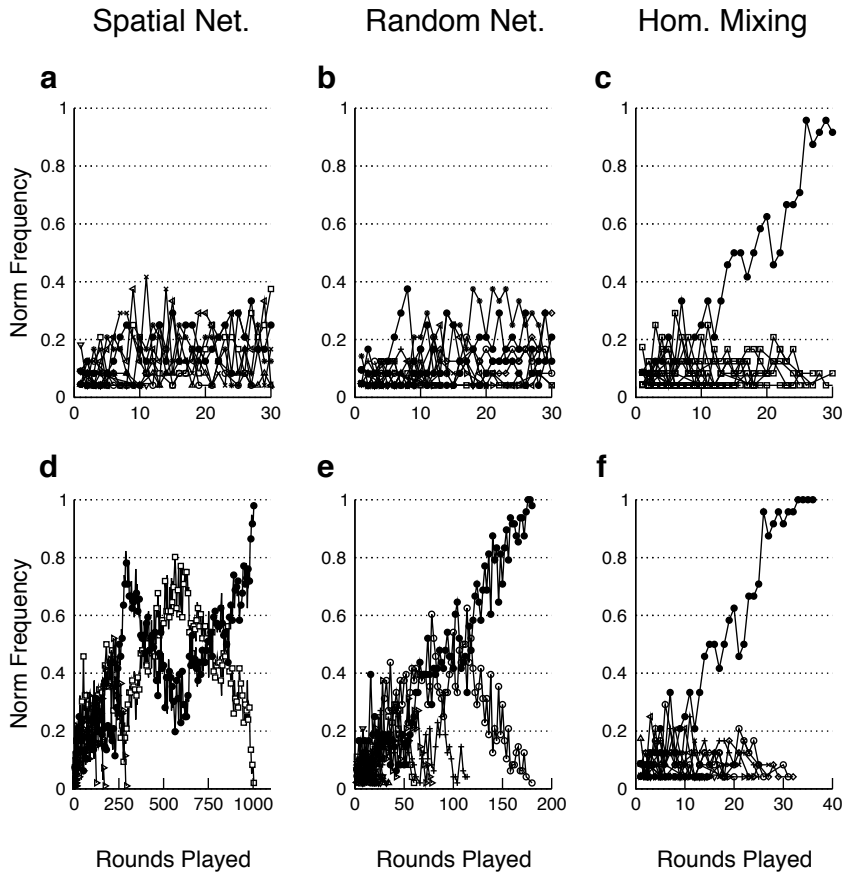

**Figure S6. Numerical simulations of evolution to final consensus. (a,d)** Spatial network, **(b,e)** Random network, **(c,f)** homogeneously mixing population. The top panels **(a-c)** show the temporal evolution in the first 30 rounds, i.e. in the experimentally accessible regions of the dynamics. The bottom panels **(d-f)** show the same simulations run until final convergence. For the spatial network final consensus requires more than 1000 rounds of play. After approximately 300 rounds only two conventions remain in the population, and they swap their rank twice as the dynamics proceeds through local coarsening. In the random graph, the initial clusters of local coordination are more permeable due to the lack of a spatial structure and therefore permit symmetry breaking on longer timescales. In Figure, local coarsening shifts to symmetry breaking within ~120 rounds, and the population reaches convergence by 180 rounds. In the homogeneously mixing population, rapid symmetry breaking leads to convergence on accelerated timescales detectable within the experimental window.

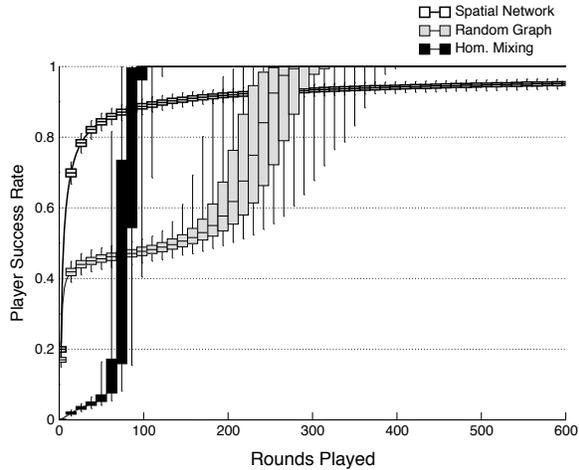

**Figure S7. Large *N* simulations of the Player Success Rate (*N*=1000).** Average success in the spatial networks (white boxes) grows rapidly in the first 50 rounds because repeated interactions within neighborhoods facilitates local coordination. However, final consensus time is protracted by local competition among emergent groups. In random networks (grey boxes), an initial phase of rapid local consensus with slow global consensus shifts toward a sudden jump of the player success rate. This is a typical signature of the underlying symmetry breaking process eventually occurring in the space of conventions. In homogeneously mixing populations (black boxes), initial coordination is most difficult during the first 50 rounds. However, this gives way to a sharp symmetry breaking transition, in which a global convention spontaneously emerges.